\documentclass[onecolumn,amsmath,amssymb,11pt,superscriptaddress,nofootinbib]{revtex4-2}
\usepackage[T1]{fontenc}
\usepackage{lmodern}
\usepackage[british]{babel}
\makeatletter\AtBeginDocument{\let\@elt\relax}\makeatother
\usepackage{amsmath}
\usepackage{amssymb}
\usepackage{physics}
\usepackage{amsthm}
\usepackage[]{graphicx}
\usepackage[justification=centering]{caption}
\usepackage{subcaption}
\usepackage{tensor}
\usepackage[dvipsnames]{xcolor}
\usepackage{cancel}
\usepackage{setspace}
\usepackage{fancyhdr}
\usepackage{appendix}
\usepackage{url}
\usepackage{natbib}
\usepackage{hyperref}
\hypersetup{
	colorlinks = true,
	linkbordercolor = {white},
	linkcolor=black,
	citecolor=blue,
	urlcolor=black}
\usepackage[a4paper,margin=2.5cm]{geometry}

\def\e{{\rm e}}
\newcommand{\nn}{\nonumber}
\begin{document}
	
	\allowdisplaybreaks
	
	\title{Stability of Axion-Dilaton Wormholes  \vspace{0.5cm}}
	
	\author{Caroline Jonas}
	\email[]{caroline.jonas@kuleuven.be}
	\affiliation{Institute for Theoretical Physics, KU Leuven,\\ Celestijnenlaan 200D, B-3001 Leuven, Belgium}
	\author{George Lavrelashvili}
	\email[]{george.lavrelashvili@tsu.ge}
	\affiliation{Department of Theoretical Physics, A.Razmadze Mathematical Institute \\
		at I.Javakhishvili Tbilisi State University, GE-0193 Tbilisi, Georgia}
	\author{Jean-Luc Lehners}
	\email[]{jlehners@aei.mpg.de}
	\affiliation{Max Planck Institute for Gravitational Physics \\ (Albert Einstein Institute), 14476 Potsdam, Germany}
	
	\begin{abstract}
		\vspace{0.5cm}
		We study the perturbative stability of Euclidean axion-dilaton wormholes that asymptotically approach flat space, both with a massless and a massive dilaton, and focussing on homogeneous perturbations. We find massless wormholes to always be perturbatively stable. The phenomenologically more relevant case of a massive dilaton presents us with a wide variety of wormhole solutions, depending on the dilaton coupling and mass, and on the axion charge. We find that the solutions with the smallest dilaton potential are perturbatively stable and dominant, even in cases where the wormhole solutions are not continuously connected to the massless case by decreasing the mass. For branches of solutions emanating from a bifurcation point, one side of the branch always contains a negative mode in its spectrum, rendering such solutions unstable. The existence of classes of perturbatively stable wormhole solutions with massive dilaton sharpens the puzzles associated with Euclidean wormholes. 
	\end{abstract}
	
	\maketitle
	
	\tableofcontents
	
	\section{Introduction}
	
Axion-dilaton wormholes \cite{Giddings:1987cg,Andriolo:2022rxc,Jonas:2023ipa} are finite action Euclidean solutions to the coupled axion-dilaton-gravity theory in four dimensions, which are expected to contribute significantly to the Euclidean path integral if they are stable.

These solutions can mediate topology change in four dimensions, leading to a variety of puzzles. More specifically, Euclidean wormholes lead to the apparent loss of unitarity \cite{Hawking:1987mz,Lavrelashvili:1987jg,Lavrelashvili:1988jj,Giddings:1988cx} which can be resolved by the introduction of so-called Coleman $\alpha$-parameters \cite{Coleman:1988cy}. These are spacetime independent quantities arising as coupling constants for the various operators in the Lagrangian. Effectively one is trading non-unitarity for the lack of knowledge of these coupling constants, leading to a multiverse type of picture. However $\alpha$-parameters stand in contradiction with quantum gravity conjectures such as the ``No global symmetry'' conjecture \cite{Vafa:2005ui,Palti:2019pca} and they are indeed absent from specific string theory settings \cite{Rey:1998yx,Maldacena:2004rf,Arkani-Hamed:2007cpn}. In parallel, Euclidean wormholes are in sharp tension with the holographic principle, since they can connect two distinct asymptotic boundaries and break the factorisation of the partition function which would be expected from the holographic viewpoint. All these problems\footnote{Detailed reviews of the many paradoxes skimmed over here are \cite{Hebecker:2018ofv,Kundu:2021nwp}, while \cite{Loges:2023ypl} presents an embedding in 10 dimensions.} call for a better understanding of wormhole solutions and in particular of their stability, as the existence of negative modes would discard their contribution to the path integral and offer an easy escape.

In their original research \cite{Giddings:1987cg} Giddings and Strominger (GS) found two types of solutions: wormholes in axion-gravity and  ``stringy'' wormholes in axion-gravity with a massless dilaton field added. In particular, it was found that the latter solutions 
only exist if the dilatonic coupling $\e^{\beta \phi}$ contains a coupling constant $\beta$  
smaller than some critical value $\beta_c$,  $0 \leq \beta < \beta_c \equiv 2 \sqrt{\frac{2}{3}}$. 

The stability analysis of axionic wormholes, i.e.,~the investigation of solutions of axion-gravity theory and its fluctuations, without extra fields,
already has some history. It was claimed in \cite{Rubakov:1996cn} that the GS wormhole has exactly one negative mode in its lowest (homogeneous) perturbation mode. This statement was disproved in \cite{Alonso:2017avz}, where it was shown that with the proper choice of (gauge-invariant) variables there are no dynamical degrees of freedom in the homogeneous perturbation sector of axion-gravity theory and therefore  
there are no homogeneous negative modes. Later it was claimed that axion-gravity wormholes have multiple instabilities in their higher angular harmonics \cite{Hertog:2018kbz} and hence the above--mentioned puzzles are eliminated as these wormholes then cannot be relevant saddle points of the Euclidean path integral.
But recently this statement was refuted in \cite{Loges:2022nuw} where, working with proper gauge invariant variables and imposing appropriate boundary conditions, 
it was demonstrated that axion-gravity wormholes are linearly stable.  

\begin{figure}[h]
		\centering
		\includegraphics[width=0.48\textwidth]{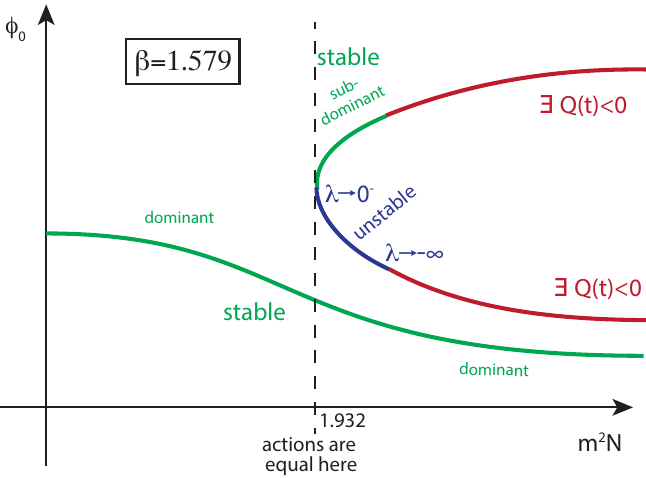}\hspace{0.5cm}
		\includegraphics[width=0.48\textwidth]{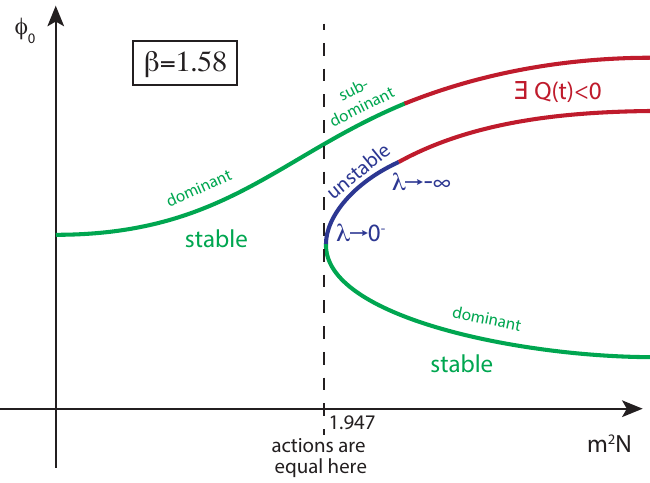}
		\caption{\small Summary of the branch structure and stability properties for axion-dilaton wormhole solutions, for dilaton coupling $\beta=1.579$ (just below hierarchy inversion, left panel) and $\beta=1.58$ (just above hierarchy inversion, right panel). The solutions are indicated by their dilaton field value $\phi_0$ at the neck, and as a function of $m^2N,$ where $m$ is the dilaton mass and $N$ the axion charge. At sufficiently large $m^2N$ new branches of solutions appear, starting from a bifurcation point. The bifurcation point admits an extra zero mode, which turns into a negative mode along one branch of solutions (shown in blue). For dilaton couplings below a critical value $\beta_i$ the lowest $\phi_0$ solutions are stable and dominant, and moreover they are continuously connected to the massless solution. Above this dilaton coupling $\beta_i,$ the branch that connects to the massless case does not remain dominant at large $m^2N,$ where a new branch of solutions takes over. More details are provided in section \ref{sec:stability}. }\label{fig:summary}
	\end{figure}

The aim of the present study is to investigate the linear stability of ``stringy'' GS wormholes 
and their generalisations with a {\it massive} dilaton field \cite{Andriolo:2022rxc,Jonas:2023ipa}\footnote{GS wormholes with a {\it massless} dilaton have independently been claimed to be perturbatively stable \cite{Maenaut:202X}.}.
In a previous  paper \cite{Jonas:2023ipa}, in particular, we found new branches of solutions when the dilaton field is massive, exhibiting an interesting bifurcating structure crucially depending on the value of the dilatonic coupling constant $\beta$. These solutions significantly enlarge the bestiary of known axionic wormholes -- examples for two representative values of the dilaton coupling $\beta$ are shown in figure \ref{fig:summary}.
Due to a scaling symmetry \cite{Jonas:2023ipa}, solutions depend on the dilaton mass $m$ and the axion charge $N$ only through the combination
$m^2 N$. It was found that for $\beta < \beta_c$ there always exists a branch which in the limit $m^2 N \to 0$ coincides with the GS stringy wormholes. On top of that a new bifurcating branch of solutions appears above some threshold value of  $m^2 N.$  
The initial values of the dilaton field (i.e.,~at the neck of the wormhole geometry) for the bifurcating branch lie above the GS branch for small $\beta$,  $\beta < \beta_i$, where    
$1.579 < \beta_i < 1.580$, see the left panel in figure \ref{fig:summary}. Above $\beta_i$ the branch structure is inverted and the 
initial field values of the dilaton for the bifurcating branch lie below the GS branch. This is illustrated in the right panel of figure \ref{fig:summary}.
 	
Euclidean field configurations contribute to the path integral via their Euclidean action $S_E,$ with a factor $\e^{-S_E/\hbar}.$ The perturbative stability is analysed as usual by considering perturbations up to quadratic order in the action:
	\begin{equation}
		S_\text{E}[\bar{x}+X]\simeq S_E[\bar{x}]+\frac{1}{2}\left.\frac{\delta ^2S_E}{\delta x^2}\right|_{\bar{x}} X^2\,,
	\end{equation}
where $\bar{x}$ is background (wormhole) solution and $X$ are small perturbations of the background.
More specifically, in its homogeneous sector we find that axion-dilaton wormholes admit a single physical gauge-invariant variable ${\cal R}$, which is a curvature perturbation that also contains fluctuations of the dilaton field. Its (unconstrained) quadratic action is of the form
	\begin{equation}
		S^{(2)}=\frac{1}{2}\int\dd^4x\sqrt{g}\left(\frac{\dot\phi^2}{Q(a, \phi)}\dot{{\cal R}}^2+U(a, \phi) {\cal R}^2\right),\label{quad}
	\end{equation}
	where a dot denotes a Euclidean time derivative. The factor $Q$ in the kinetic term as well as the potential $U$ depend on background quantities. When $Q>0,$ this corresponds to a regular Sturm-Liouville problem, with eigenvalue equation 
	\begin{equation}
		-\frac{1}{a^3}\frac{\dd}{\dd t_E}\left(a^3\frac{\dot\phi^2}{Q}\dot{{\cal R}}\right)+ U {\cal R}=\lambda w {\cal R}\,, \label{eigen}
	\end{equation}
where $\bar{x}=\{a, \phi \}$ is the background (wormhole) solution (scale factor and dilaton field profiles), $\lambda$ is an eigenvalue and $w$ a 
measure\footnote{Note that whereas in analysing the linear stability of static solutions the measure automatically follows from the action, 
for the stability analysis of Euclidean solutions we are free to choose a reasonable measure. The choice of (nonsingular) measure affects the numerical  values of eigenvalues, but does not change the number of negative modes of the corresponding Sturm-Liouville problem.}.
When the eigenvalues are all positive, then any small perturbation will increase the value of the Euclidean action and thus decrease the weighting of the solution. In such cases, the wormholes are (at least perturbatively) stable. Meanwhile, when a negative eigenvalue exists, it indicates that the solution can be deformed into a lower action solution with higher weighting. This signals an instability and that the wormhole will be sub-dominant compared to the related solution without negative mode, which we find to exist at the same parameter values in all cases.

The diagrams in figure \ref{fig:summary} summarise our main findings for two representative branch structures. The curves shown in green denote solutions that are stable, i.e.,~for which all eigenvalues are positive. For some of these solutions both the kinetic factor $Q$ and the potential $U$ are positive throughout, so that stability is obvious. In particular, we find that the branch with the lowest initial dilaton values is always perturbatively stable. For $\beta<\beta_i$ this branch continuously connects to the massless solutions as $m^2N$ is lowered. However, for $\beta>\beta_i$ there is a hierarchy inversion and a corresponding jump in the initial dilaton value at a critical value of $m^2N$ where a bifurcation point appears, see the right panel in figure \ref{fig:summary}. In that case it is still true that the solutions with the lowest $\phi_0$ value are stable, but the solutions do not continuously connect from large to small $m^2N$ values. Note that in both of the cases shown in the figure, there are parameter regions where more than one stable solution exists. In such cases we find that the solutions with the lowest $\phi_0$ values are the most dominant.

At the bifurcation points just mentioned, an extra zero mode appears. Along one side of the associated new branch, this mode corresponds to a stable deformation of the solutions. However, on the other side of the bifurcation point the zero mode turns into a negative mode. The corresponding wormholes are unstable -- these solutions are shown in blue in the figure. Such wormholes would thus not contribute significantly to the Euclidean path integral. 

Finally, for the solutions marked in red in the figure, the kinetic factor $Q$ becomes negative somewhere along the wormhole. This implies that we do not have a regular Sturm-Liouville problem anymore, and we cannot analyse the perturbative stability in the same manner. In fact, we will not be able to say anything conclusive about these solutions but, as we will discuss in section \ref{sec:stability}, this case presents a number of similarities with certain classes of Coleman-De Luccia bounces. 

For values of the dilaton coupling above the critical GS value $\beta>\beta_c,$ no massless wormholes exist. In these cases only the bifurcating branches exist, at sufficiently large $m^2N$ \cite{Jonas:2023ipa}. Their stability properties are entirely analogous to those of the bifurcating branches that arise for $\beta>\beta_i,$ i.e. they resemble the bifurcating branch shown in the right panel of the figure (with the branch connecting to $m^2N=0$ simply removed). 
	
Finally we will also show that the other type of axion-dilaton wormholes studied in \cite{Jonas:2023ipa} --  wormholes leading to the creation of expanding baby universes (\textit{expanding} wormholes for short) -- necessarily have $Q<0$ somewhere along the solution, and hence we cannot study their perturbative stability reliably at this point.

The main conclusion of our work is that large classes of massless and massive axion-dilaton wormholes are stable, and that the puzzles mentioned at the beginning of this section are indeed present and await resolution. These puzzles provide a clear opportunity for gaining new insights into quantum gravity.
	
The rest of the paper is organised as follows: in section \ref{sec:bckg} we briefly review the generalised Giddings-Strominger wormhole solutions found in \cite{Jonas:2023ipa}. Then in section \ref{sec:pert} we compute the unconstrained quadratic action in the Lagrangian approach and obtain the relevant fluctuation equation. In section \ref{sec:stability}, we solve the fluctuation equation for concrete numerical background solutions both with massless and massive dilaton, from which we deduce their perturbative (in)stability. The alternative Hamiltonian approach to the fluctuation equation is presented in Appendix \ref{app:hamiltonian}, where we also discuss why this approach fails in the present context. Our conclusions are drawn in section \ref{sec:conclusions}.
	
	
	\section{Wormhole solutions}\label{sec:bckg}
	
We consider the (Lorentzian) axion-dilaton-gravity action
	\begin{equation}
		S_\text{L}=\int\dd^4x\sqrt{-g}\left[\frac{1}{2\kappa^2}R-\frac{1}{2}\nabla_{\mu}\phi \nabla^{\mu}\phi
		-V(\phi)-\frac{e^{-\beta \phi\kappa}}{12 f^2}  H_{\mu\nu\rho}H^{\mu\nu\rho} \right],\label{eq:fullactionlorentzian}
	\end{equation}
	where $\kappa^2\equiv M_\text{Pl}^{-2}=8\pi G$, $\beta$ is the dilatonic coupling, the dilaton potential is $V(\phi)$ and $H_{\mu\nu\rho}$ is the 3-form field strength of an axion field with coupling $f$. We will only consider a potential that provides a mass to the dilaton,
	\begin{align}
	V(\phi) = \frac{1}{2} m^2 \phi^2\,.
	\end{align}
	The equations of motions are
	\begin{equation}
		\left\lbrace
		\begin{aligned}
			&R_{\mu\nu}=\kappa^2 \partial_\mu \phi\partial_\nu \phi + \kappa^2 V g_{\mu\nu} +\frac{\kappa^2}{2 f^2} \e^{-\beta \phi\kappa} H_{\mu\rho\sigma} H_\nu^{\rho\sigma}-\frac{\kappa^2}{6 f^2}\e^{-\beta \phi\kappa} H_{\gamma\rho\sigma}H^{\gamma\rho\sigma} g_{\mu\nu}\,, \\
			&\frac{1}{\sqrt{g}}\partial_\mu(\sqrt{g}g^{\mu\nu} \partial_\nu \phi)= \frac{\partial V}{\partial\phi}
			- \frac{\beta\kappa}{12 f^2} 		\e^{-\beta \phi\kappa} H_{\gamma\rho\sigma}H^{\gamma\rho\sigma}\,,\\
			&\partial_\mu (\sqrt{g} \e^{-\beta\phi\kappa} H^{\mu\rho\sigma})= 0 \,.
		\end{aligned}
		\right.\label{eq:eomfullaxiongravitydilaton}
	\end{equation}
The axion field strength $H=\dd B$ is the exterior derivative of a 2-form, hence the Bianchi identity holds:
	\begin{equation}
		\dd H=0\ \Leftrightarrow\ \nabla_{[\mu}H_{\nu\rho\sigma]}=0\,.\label{eq:BianchiID}
	\end{equation}
	We take a homogeneous and isotropic ansatz in (Lorentzian) conformal time $\eta$,
	\begin{equation}
		\left\lbrace
		\begin{aligned}
			&\dd s^2=a^2(\eta)(-\dd \eta^2+\dd\Omega_3^2)\,,\\
			&\phi=\phi(\eta)\,,\\
			&H_{0ij}=0\,,\quad H_{ijk}=q\,\varepsilon^{\text{N}}_{ijk}=\frac{q}{\sqrt{g^{(3)}}}\varepsilon^{\text{T}}_{ijk}\,,
		\end{aligned}
		\right.\label{eq:conformalansatz}
	\end{equation}
where $\dd\Omega_3^2$ is the metric on the round $3$-sphere.	This ansatz automatically satisfies the Bianchi identity \eqref{eq:BianchiID}. In what follows, primes will denote derivatives with respect to $\eta$. $g^{(3)}$ is the determinant of the induced 3-dimensional metric on constant $\eta$ hypersurfaces. $\varepsilon^{\text{N}}$ is the tensor density taking numerical values $\lbrace\pm1,0\rbrace$, while $\varepsilon^{\text{T}}$ is the tensor so that the volume of a spatial sphere is
	\begin{equation}
		\int_{S^3}\varepsilon^T=2\pi^2a^3\,.
	\end{equation}
	Therefore the flux of the 3-form field through the same spatial sphere is
	\begin{equation}
		\int_{S^3}H=2\pi^2q\,.
	\end{equation}
	For convenience we redefine the charge parameter of the wormhole solutions as $\displaystyle N^2\equiv\frac{q^2}{2f^2}$. Plugging \eqref{eq:conformalansatz} into the action \eqref{eq:fullactionlorentzian} and equations of motion \eqref{eq:eomfullaxiongravitydilaton}, we obtain
	\begin{align}
		S_\text{L}&=2\pi^2\int d\eta \left( \frac{1}{\kappa^2}(3a^2 + 3 a a'') + \frac{1}{2}a^2 \phi'^2 - a^4 V -e^{-\beta\phi\kappa}\frac{N^2}{a^2}\right) \,,
	\end{align}
	and
	\begin{align}
			0 &= 2 \frac{a''}{\kappa^2 a} -\frac{a'^2}{\kappa^2 a^2} + \frac{1}{\kappa^2} + \frac{1}{2}\phi'^2 -a^2 V + \frac{N^2}{a^4}e^{-\beta\phi\kappa}\,, \label{eoma}\\
			0 &=  3\frac{a'^2}{\kappa^2 a^2} + \frac{3}{\kappa^2} - \frac{1}{2}\phi'^2 - a^2 V - \frac{N^2}{a^4}e^{-\beta\phi\kappa}\,, \label{constraint}\\
			0 &= \phi'' + 2 \frac{a'}{a}\phi' + a^2 V_{,\phi} - \frac{\beta \kappa N^2}{a^4}e^{-\beta\phi\kappa}\,. \label{eomphi}
	\end{align}
	The on-shell action simplifies to
	\begin{equation}
		S^\text{on-shell}_L=2\pi^2\int\dd\eta\left(a^4V(\phi)- 2 e^{-\beta\phi\kappa}\frac{N^2}{a^2}\right).
	\end{equation}

	The above equations have been worked out in Lorentzian conformal time to ease the calculations of perturbations in next section, but it is also useful to transform them to Euclidean non-conformal time ($\dd t_\text{E}=-ia(\eta)\dd\eta$), in which variable the numerical solutions in \cite{Jonas:2023ipa} were found. The equations of motion then read
	\begin{equation}
		\textrm{Euclidean}\qquad \left\lbrace
	\begin{aligned}
		 0 &= \frac{\dot{H}}{\kappa^2} + \frac{1}{2}\dot\phi^2 +\frac{1}{\kappa^2a^2} - \frac{N^2}{a^6}\e^{-\beta\phi\kappa}\,,\\
		 0 &=  \frac{3H^2}{\kappa^2} - \frac{3}{\kappa^2a^2} - \frac{1}{2}\dot\phi^2 + V + \frac{N^2}{a^6}\e^{-\beta\phi\kappa}\,, \\
		 0 &= \ddot\phi + 3H\dot\phi - V_{,\phi} + \frac{\beta\kappa N^2}{a^6}\e^{-\beta\phi\kappa}\,, 
	\end{aligned}
	\right.\label{eomE}
	\end{equation}
	where dots denote derivatives with respect to $t_\text{E}$ and $H=\dot{a}/a$.
	
	\begin{figure}[h!]
		\includegraphics[width=0.75\linewidth]{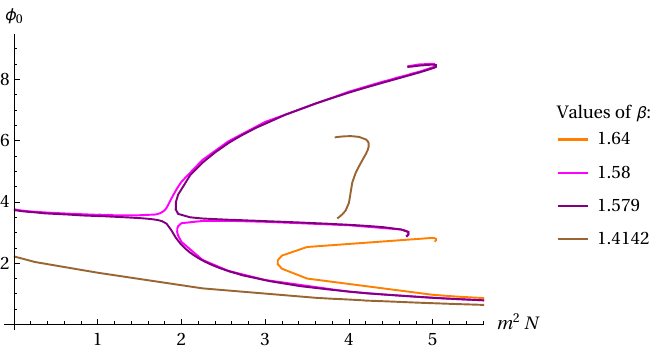}{\small\caption{Existence of axion-dilaton wormhole solutions, as found in \cite{Jonas:2023ipa} for four typical values of $\beta$. The solutions are displayed via their initial dilaton value $\phi_0,$ as a function of their mass and charge combination $m^2N.$ Below the critical value $\beta_c=2\sqrt{\frac{2}{3}}\approx 1.63$ solutions exist which are continuously connected to the massless case. In addition, for all dilaton couplings new bifurcating branches appear at sufficiently large values of $m^2N.$ Full details can be found in \cite{Jonas:2023ipa}.\label{fig:bckgsol}}}
	\end{figure}
	
\begin{figure}[h!]
		\includegraphics[width=0.45\linewidth]{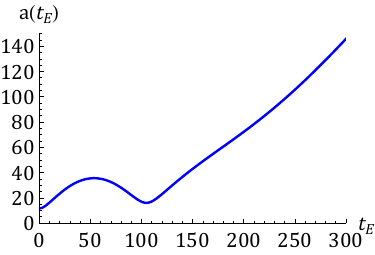}\qquad
		\includegraphics[width=0.45\linewidth]{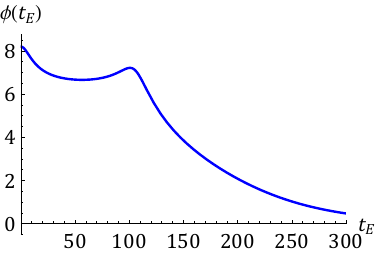}
		{\small\caption{An example of a Euclidean wormhole solution displaying oscillations in the fields. The evolution of the scale factor is shown on the left (linear growth at late times indicates flat space in polar coordinates), and that of the scalar field on the right. We are only plotting the evolution from the neck of the wormhole to infinity, as would be appropriate for describing the nucleation of a baby universe. The full wormhole connecting two asymptotically flat regions would then be obtained by adding a reflection across $t_E=0.$ Here $\beta=1.2,\, m^2N=3,\, \phi_0=8.1964321797$.\label{fig:example}}}
	\end{figure}

	Euclidean wormhole solutions are found by numerically solving the above system of equations with initial boundary conditions $\dot{a}(t_\text{E}=0)=0,\,\dot{\phi}(t_\text{E}=0)=0$ (our action did not include a surface term and thus these Neumann type boundary conditions are appropriate). The Friedmann constraint \eqref{constraint} provides a relation between the initial field values $a_0$ and $\phi_0$:
	\begin{equation}
		\frac{3}{\kappa^2a_0^2}=V(\phi_0)+\frac{N^2\e^{-\beta\kappa\phi_0}}{a_0^6}\,. \label{initialvalues}
	\end{equation}	
	Actual solutions are then found by tuning $\phi_0,$ say, such that they approach flat space asymptotically. The unexpected result from \cite{Jonas:2023ipa} is that for some ranges of the parameters $\beta, m^2N$ of the theory, solutions with different initial dilaton field values $\phi_0$ coexist, see figure \ref{fig:bckgsol}. All of these solutions start at a minimum scale factor value (the ``neck'' of the wormhole) and asymptote to flat space. However, some solutions develop additional features, such as oscillations in the dilaton field and also additional extrema in the scale factor. As a general rule, the higher the initial value $\phi_0,$ the higher the complexity of the field evolution -- see figure \ref{fig:example} for an example of a wormhole solution with additional oscillations (more illustrations are provided in \cite{Jonas:2023ipa}). In what follows, we would like to determine the perturbative stability of these solutions.
	
	\section{Linear Perturbations}\label{sec:pert}

	We are now ready to study perturbations around the wormhole solutions that we just described. We introduce metric perturbations $A,\,\Psi$ in the conformal Newtonian gauge and matter perturbations $\Phi,\, W,\, S$ on top of the background \eqref{eq:conformalansatz}
	\begin{equation}
		\left\lbrace
		\begin{aligned}
			&\dd s^2=a^2(\eta) \left[- (1+2A(\eta,x^i)) \, \dd \eta^2 +  (1-2\Psi(\eta,x^i)) \, \dd\Omega_3^2\right],\\
			&\phi=\phi(\eta)+\Phi(\eta,x^i)\,,\\
			&H_{0ij}= 0 + \varepsilon^N_{ijk} \delta^{kl}\partial_l W(\eta,x^i)= \frac{q}{a}\varepsilon^T_{ijk}g^{kl}\partial_l W(\eta,x^i)\,,\\ &H_{ijk}=q \varepsilon^N_{ijk}\left( 1+S(\eta,x^i)\right) =\frac{q}{\sqrt{g^{(3)}}} \varepsilon^T_{ijk}\left( 1+S(\eta,x^i)\right).
		\end{aligned}
		\right.\label{eq:sphericalansatz}
	\end{equation}
	Plugging this ansatz into the Lorentzian action, and extracting the part quadratic in the perturbations, we obtain (using the background equations of motion and integrations by parts):
	\begin{align}
		S^{(2)}_\text{L} =  \int d^4x \sqrt{\gamma}  a^2 & \Bigg[ A^2 \bigg(  \frac{3}{\kappa^2} -a^2 V - \frac{N^2}{a^4}e^{-\beta\phi\kappa}\bigg)- 3\Psi^2 \bigg(\frac{1}{\kappa^2}  + \frac{3 N^2}{a^4}e^{-\beta\phi\kappa} \bigg) \nn \\ & \left.
		+ 6A\Psi \left( \frac{1}{\kappa^2} - \frac{N^2}{a^4}e^{-\beta\phi\kappa}\right) +\frac{1}{\kappa^2}(2A-\Psi)\nabla^2\Psi- \frac{3}{\kappa^2} \Psi'^2 -6\frac{a'}{a\kappa^2 }A\Psi' \right. \nn \\ & \left.
		-\Phi^2 \bigg(\frac{1}{2}a^2 V_{,\phi\phi}+ \frac{\beta^2\kappa^2}{2}\frac{N^2}{a^4}e^{-\beta\phi\kappa}\bigg) - A\Phi \bigg(a^2 V_{,\phi} -\beta\kappa\frac{N^2}{a^4}e^{-\beta\phi\kappa}\bigg)    \right. \nn \\ & \left.
		+ \frac{1}{2}\Phi'^2 + \frac{1}{2}\Phi\nabla^2\Phi-\phi' A \Phi'+ 6\beta\kappa\frac{N^2}{a^4}e^{-\beta\phi\kappa}\Psi\Phi + 3\phi'  \Psi' \Phi
		 \right. \nn \\ &
		-\frac{N^2}{a^4}e^{-\beta\phi\kappa}\left(S^2 + W\nabla^2 W\right) - \frac{2N^2}{a^4}e^{-\beta\phi\kappa}S\left(A+3\Psi-\beta\kappa\Phi\right)\Bigg]. \label{eqac}
	\end{align}
	In the above, $\sqrt{\gamma}=\sin^2\theta \sin\phi$ is the determinant of the metric on the unit $3-$sphere and $\nabla$ the associated covariant derivative. Our calculational strategy here is to consider perturbations on the full wormhole spacetime, with two asymptotically flat regions and the wormhole neck in the middle. This allows us to disregard surface terms until a later stage -- we will comment further on this point below. The expression \eqref{eqac} reduces to the actions found previously in \cite{Khvedelidze:2000cp,Gratton:1999ya,Gratton:2000fj} when the axion is turned off and in \cite{Loges:2022nuw} when the dilaton is turned off. 
	
The canonically conjugate momenta are given by
	\begin{align}
		\Pi_\Psi & = a^2\sqrt{\gamma}\left( -\frac{6}{\kappa^2}\Psi' - \frac{6}{\kappa^2} \frac{a'}{a} A + 3 \phi' \Phi \right),\\
		\Pi_{\Phi} & = a^2 \sqrt{\gamma} \left( \Phi' -\phi' A\right),\\
		\Pi_A &=0\,, \quad \Pi_S = 0\,, \quad \Pi_W=0\,.
	\end{align}
	Under a gauge transformation $\eta\to \eta+\lambda,$ the perturbations transform as
	\begin{align}
		A \to A + \lambda' + \frac{a'}{a}\lambda\,, \quad & \Psi \to \Psi - \frac{a'}{a}\lambda\,, \quad \Phi \to \Phi + \phi' \lambda\,, \label{eq:gauge}\\
		\Pi_\Psi \to \Pi_\Psi + 6a^2\sqrt{\gamma}\left(\frac{1}{\kappa^2}-\frac{N^2}{a^4}e^{-\beta\phi\kappa}\right)\lambda\,, \quad & \Pi_\Phi \to \Pi_\Phi + a^2\sqrt{\gamma}\left(\phi''-\frac{a'}{a}\phi'\right)\lambda\,.
	\end{align}
	Moreover under a ``spatial'' diffeomorphism, $S$ and $W$ transform but $S'-\nabla^2W$ is gauge invariant \cite{Loges:2022nuw}. In fact, the linearised Bianchi identity sets this combination to zero, $S'-\nabla^2W=0.$ Then, in the homogeneous sector, $S$ is constant and thus zero if we fix the charge asymptotically. Hence, in the homogeneous sector, the $S$ and $W$ perturbations will play no role.
	
	In the following we restrict to the homogeneous mode ($\nabla\Psi = 0 = \nabla \Phi$) because inhomogeneous modes add to the gradient energy, hence if the homogeneous mode is stable, the inhomogeneous ones will typically not spoil this stability. As argued above, in the homogeneous sector the axionic perturbations are not dynamical: $S=0=W,$ and the action \eqref{eqac} becomes
	\begin{align}
		S^{(2)}_\text{L} =  \int d^4x \sqrt{\gamma}  a^2  \Bigg[& -\frac{3}{\kappa^2} {\mathcal{Q}} A^2
		-3 \Psi^2 \bigg(\frac{1}{\kappa^2}  +\frac{3 N^2}{a^4}e^{-\beta\phi\kappa} \bigg) - \frac{3}{\kappa^2} \Psi'^2 + 6A\Psi \bigg( \frac{1}{\kappa^2} - \frac{N^2}{a^4}e^{-\beta\phi\kappa}\bigg)  \nn \\ & \left.
		-\Phi^2 \bigg(\frac{1}{2}a^2 V_{,\phi\phi}+ \frac{\beta^2\kappa^2}{2}\frac{N^2}{a^4}e^{-\beta\phi\kappa}\bigg) -A\Phi \bigg( a^2 V_{,\phi} - \beta\kappa\frac{N^2}{a^4}e^{-\beta\phi\kappa}\bigg)+ \frac{1}{2}\Phi'^2
		 \right. \nn \\ &
		 -6\frac{a'}{a\kappa^2 }A\Psi'-\phi' A \Phi'
		+ 6\Psi\Phi\beta\kappa\frac{N^2}{a^4}e^{-\beta\phi\kappa} + 3\phi' \Psi' \Phi \Bigg],
	\end{align}
	where the conformal Hubble rate is ${\cal H}=a'/a$ and we have defined
	\begin{align}
		{\mathcal{Q}} \equiv {\cal H}^2 - \frac{\kappa^2}{6}\phi'^2\,.
	\end{align}
	We now rewrite this action in terms of gauge-invariant variables by using the gauge-invariant lapse ${\cal A}$ and curvature perturbation ${\cal R}$ (see equation \eqref{eq:gauge}),
	\begin{align}
		{\cal A} & = A + \frac{1}{\cal H}\Psi' + \left( \frac{{\cal H}'}{{\cal H}\phi'} - \frac{{\cal H}}{\phi'}\right) \Phi \\
		{\cal R} & = \Psi + \frac{{\cal H}}{\phi'} \Phi.
	\end{align}
	We obtain the manifestly gauge-invariant expression
	\begin{align}
		S^{(2)}_\text{L} =  \int d^4x \sqrt{\gamma}  a^2 & \left[-\frac{3}{\kappa^2}{\cal Q}{\cal A}^2 + \frac{\phi'^2}{2{\cal H}^2}{\cal R}'^2 - \frac{\phi'^2}{{\cal H}} {\cal A}{\cal R}' + 6\left(\frac{1}{\kappa^2}-\frac{N^2}{a^4}e^{-\beta\kappa\phi}\right){\cal A}{\cal R}  \right. \nonumber \\ & \left. \quad + \left(3\beta\kappa\frac{\phi'}{{\cal H}}\frac{N^2}{a^4}e^{-\beta\kappa\phi}- \frac{3}{{\cal H}^2}\Big( a^2\kappa^2V-3{\cal H}^2-2\Big)\Big( \frac{1}{\kappa^2}-\frac{N^2}{a^4}e^{-\beta\kappa\phi}\Big)\right) {\cal R}^2 \right].
	\end{align}
	We can further eliminate the gauge-invariant lapse ${\cal A}$, as it is not a propagating degree of freedom but rather an auxiliary field, using its (constraint) equation
	\begin{align}
		\frac{1}{\kappa^2}{\cal Q}{\cal A} =  - \frac{\phi'^2}{6{\cal H}} {\cal R}' + \left(\frac{1}{\kappa^2}-\frac{N^2}{a^4}e^{-\beta\kappa\phi}\right){\cal R} \,,
	\end{align}
	to end up with 
	\begin{align}
		S^{(2)}_\text{L} =\frac{1}{2} \int d^4x \sqrt{\gamma} a^2 & \Bigg\{\frac{\phi'^2}{{\mathcal{Q}}}{\cal R}'^2  + \frac{2\kappa^2}{{\cal Q}^2}\bigg[ -\frac{a^2\kappa^2V \phi'^2}{3}\Big(\frac{1}{\kappa^2}-3\frac{N^2}{a^4}e^{-\beta\kappa\phi}\Big) -a^2V_{,\phi} \phi' {\cal H}\Big(\frac{1}{\kappa^2}-\frac{N^2}{a^4}e^{-\beta\kappa\phi}\Big)  \nonumber \\  & \qquad\qquad \qquad \qquad- \Big(\frac{4}{3}\phi'^2 - \frac{\beta}{\kappa} {\cal H}\phi'(a^2\kappa^2V-2)\Big)\frac{N^2}{a^4}e^{-\beta\kappa\phi}\bigg]{\cal R}^2 \Bigg\}.
	\end{align}
	This is the gauge-invariant quadratic part of the action in the Lorentzian conformal time. 
	
	We can now return to Euclidean physical time $\dd t_\text{E}=-ia(\eta)\dd\eta$ and obtain
	\begin{align}
		S^{(2)} _\text{E}
		= \frac{1}{2}\int_0^\infty dt_\text{E} \int d^3x \sqrt{\gamma}  & a^3 \left(\frac{\dot\phi^2}{{Q}}\dot{\cal R}^2  + U{\cal R}^2 \right) - \int d^3x \sqrt\gamma a^3 \frac{\dot\phi^2}{{Q}}\left.{\cal R}\dot{\cal R}\right|_{t_\text{E}=0}^{t_\text{E}=+\infty}\,, \label{eq:S2inphysicaltime}
	\end{align}
	where we explicitly split up the spacetime at the location of the wormhole neck $t_\text{E}=0.$ The surface term is added in order to obtain a consistent variational problem allowing us to impose a Neumann boundary condition on ${\cal R},$ in agreement with the treatment of the background. (An equal and opposite boundary term would be added on the other half of the spacetime.) Above, we have defined
	\begin{align}
		& H\equiv\dot{a}/a={\cal H}/a,\ Q\equiv H^2-\kappa^2\dot{\phi}^2/6=\mathcal{Q}/a^2,\ {\cal R}=\Psi+\frac{H}{\dot{\phi}}\Phi=\Psi+\frac{\cal H}{\phi'}\Phi\quad \text{and} \\
		&  U=\frac{-2\kappa^2}{a^2{Q}^2}\Bigg[\frac{\kappa^2V \dot\phi^2}{3}\bigg(\frac{1}{\kappa^2}-3\frac{N^2}{a^4}e^{-\beta\phi}\bigg) +V_{,\phi} \dot\phi H\bigg(\frac{1}{\kappa^2}-\frac{N^2}{a^4}e^{-\beta\phi}\bigg)\nonumber\\
		&\qquad\qquad\qquad+\bigg( \frac{4}{3a^2}\dot\phi^2 - \frac{\beta}{\kappa} H\dot\phi\Big(\kappa^2V-\frac{2}{a^2}\Big)\bigg)\frac{N^2}{a^4}e^{-\beta\phi}\Bigg]\\
		&\qquad\ =\frac{2}{a^2Q^2}\Bigg[\frac{\kappa^2 \dot{\phi}^4}{3}-2H^2\dot{\phi}^2-\frac{\kappa^2 a^2\dot{\phi}^4}{2}\left(\dot{H}+3H^2\right)+a^2\dot{\phi}^2\left(9H^4-\dot{H}^2\right)\nonumber\\
		& \qquad\qquad \quad-3a^2H\dot{\phi}V_{,\phi}Q+a^2H\dot{\phi}\ddot{\phi}\left(\dot{H}+3H^2\right)\Bigg], \label{eq:ULag}
	\end{align}
	where the last expression was obtained using the background equations of motion.
	In physical Euclidean time, and for a weight function $w(t_\text{E}),$ the eigenvalue equation is therefore
	\begin{align}
		& -\frac{1}{a^3}\frac{\mathrm{d}}{\mathrm{d}t_\text{E}}\left( a^3 \frac{\dot\phi^2}{{ Q}} \dot{\cal R}\right) + U {\cal R} = \lambda w(t_\text{E}) {\cal R}\,,\label{eq:eigenvalueeqt}\\
		\Leftrightarrow\quad& \ddot{\cal R} + \left( 3H+2\frac{\ddot\phi}{\dot\phi} - \frac{\dot{Q}}{Q}\right) \dot{\cal R} - \left( \frac{UQ}{\dot\phi^2}-\lambda \frac{w Q}{\dot\phi^2}\right) {\cal R} = 0\,.
	\end{align}
	Before solving this eigenvalue equation on the numerical background solutions of \cite{Jonas:2023ipa}, we study its asymptotic structure both near the neck and at infinity.
	
	In the large distance limit $t_\text{E} \to \infty$, the background solutions behave as \cite{Jonas:2023ipa}:
	\begin{align}
		a \approx t_\text{E}\,, \qquad \phi(t_\text{E}) \approx \frac{\beta N^2}{m^2 t_\text{E}^6}\,,
	\end{align}
	from which we can deduce the leading behaviour of the functions appearing in the eigenvalue equation \eqref{eq:eigenvalueeqt}:
	\begin{align}
		& H \approx \frac{1}{t_\text{E}}\,, \quad \dot\phi \approx -6\frac{\beta N^2}{m^2 t_\text{E}^7}\,, \quad Q \approx \frac{1}{t_\text{E}^2} \,, \quad \frac{\dot\phi^2}{Q} \approx \frac{36\beta^2 N^4}{m^4 t_\text{E}^{12}}\,, \\ & U\approx \frac{2H\dot\phi}{a^2Q^2}\left( -V_{,\phi} -2\beta \frac{N^2}{a^6}\right) \approx \frac{36 \beta^2 N^4}{m^2 t_\text{E}^{12}}
	\end{align}
	Interestingly, the kinetic and potential terms tend to the same limit (up to a factor of $m^2$), and both are asymptotically positive. Therefore, in order to obtain well-behaved eigenfunctions, a natural choice for the weight function is the kinetic function $w=\dot\phi^2/Q$, in which case the eigenvalue equation becomes
	\begin{align}
		-\ddot{\cal R} + \frac{9}{t_\text{E}} \dot{\cal R} + (m^2-\lambda) {\cal R} \approx 0\,.
	\end{align}
	The leading solutions at large $t_\text{E}$ then asymptote to
	\begin{align}
		{\cal R} \to \e^{\pm\sqrt{m^2-\lambda}\,t_\text{E}}\,,
	\end{align}
	where we must choose the decaying branch. At infinity we then satisfy the boundary condition $\dot{\cal R}(t_\text{E} \to \infty)=0$ and moreover the boundary term in \eqref{eq:S2inphysicaltime} vanishes.
	
	At the neck of the wormhole solution at $t_\text{E}=0$, the following expansion of the fields satisfies the background equations of motion at second order in $t_\text{E},$
	\begin{align}
		a(t_\text{E}) & = a_0 + \left(\frac{1}{a_0}-\frac{\kappa^2}{4}a_0m^2\phi_0^2\right) t_\text{E}^2 + {\cal O}(t_\text{E}^4) \,,\\
		\phi(t_\text{E}) & = \phi_0+\frac{1}{2}\left(m^2\phi_0-\beta \kappa \frac{N^2}{a_0^6}e^{-\beta\kappa\phi_0}\right) t_\text{E}^2 + {\cal O}(t_\text{E}^4) \,, \\ Q(t_\text{E}) & = \left[ \left( \frac{\kappa^2 N^2}{a_0^6}\e^{-\beta\kappa\phi_0} - \frac{1}{a_0^2}\right)^2 - \frac{\kappa^2}{6}\left(  \beta \kappa\frac{N^2}{a_0^6}\e^{-\beta\kappa\phi_0} - m^2\phi_0\right)^2\right] t_\text{E}^2 + {\cal O}(t_\text{E}^4) ) \,. 
	\end{align}
	In these expressions the initial field values $a_0$ and $\phi_0$ are related by the Friedmann constraint~\eqref{initialvalues}.
	From the expansions we see that the kinetic term $\dot{\phi^2}/Q$ is regular at the origin (i.e., it can be expanded as $c_0+c_2t_\text{E}^2+\dots$). The effective potential $U$ is also regular at the origin: from its expression \eqref{eq:ULag}, the only terms for which this is nontrivial are
	\begin{align}
		\frac{2}{Q^2}\left( H\dot{H}\dot\phi\ddot\phi - \dot{H}^2 \dot\phi^2\right) & = \frac{2H^2\dot{H}\dot\phi}{Q^2}\, \frac{\dd}{\dd t_\text{E}}\left( \frac{\dot\phi}{H}\right).\label{eq:dangerousU}
	\end{align}
	Since both $\dot\phi$ and $H$ are odd functions, their ratio is even so its derivative is ${\cal O}(t_\text{E}).$ Combined with the factor $H^2\dot\phi = {\cal O}(t_\text{E}^3),$ this implies that the factor $Q^2 = {\cal O}(t_\text{E}^4)$ is compensated and the combination \eqref{eq:dangerousU} is regular. Therefore the eigenvalue equation \eqref{eq:eigenvalueeqt} is regular at the origin. It takes the form
	\begin{align}
		-\ddot{\cal R} + c_1 t_\text{E} \dot{\cal R} + (c_2 -\lambda) {\cal R} \approx 0\,,
	\end{align}
	for certain constants $c_{1,2}$, and its solution is given by the series
	\begin{align}
		{\cal R}(t_\text{E}) = {\cal R}_0 + {\cal R}_1 t_\text{E} + \frac{(c_2 - \lambda){\cal R}_0}{2}t_\text{E}^2 + \frac{(c_1+c_2-\lambda ){\cal R}_1}{6}t_\text{E}^3 + {\cal O}(t_\text{E}^4).
	\end{align}
	Both the initial field value ${\cal R}_0$ and the initial first derivative ${\cal R}_1$ are arbitrary. Since the equation is linear, we may eliminate one parameter by re-scaling. The second free parameter is fixed by considering the boundary condition at $t_\text{E}=0.$ For the background, we fixed Neumann boundary conditions $\dot{a}(0)=\dot\phi(0)=0.$ This means that we should also impose $\dot{\cal R}(0)=0,$ so that these boundary conditions are preserved. A Neumann boundary condition of this type is consistent with the variational problem obtained from equation \eqref{eq:S2inphysicaltime} and moreover we can see by inspection that the surface term will be zero at $t_\text{E}=0$ (and, incidentally, also as $t_\text{E} \to \infty$). 
	
	To summarise, we will thus focus our attention on modes that are even in Euclidean time (or, if one prefers, symmetric on the full wormhole spacetime), with boundary conditions
	\begin{align}
	\dot{\cal R}(0)=0\,, \quad {\cal R}(0)=1\,, \quad \dot{\cal R}(\infty) \to 0\,, \quad {\cal R}(\infty) \to 0
	\end{align}
We may anticipate that eigenmodes satisfying these boundary conditions will only exist for discrete eigenvalues $\lambda,$ which will need to be determined by numerical optimisation.


	\section{Stability analysis}\label{sec:stability}
	
	Using the results from the previous section we can now analyse the stability of specific wormhole solutions. We first consider the analytically known massless dilaton solutions, before focussing on the massive dilaton case. Then we comment on the so-called ``\textit{expanding}'' wormholes that we studied in \cite{Jonas:2023ipa}. Note that in all cases we study wormhole solutions from the neck out to infinity. This is the appropriate setting for describing the nucleation of baby universes. If one is interested instead in wormhole solutions linking two asymptotically flat regions, then one should add a part obtained by reflection across $t_\text{E}=0,$ which can be done trivially given the boundary conditions $\dot{a}(0)=0=\dot\phi(0).$
	
	\subsection{Massless dilaton}
	
\begin{figure}[h]
		\centering
		\includegraphics[width=0.4\textwidth]{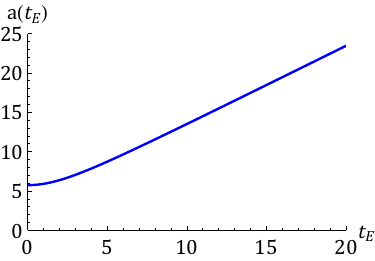}
		\includegraphics[width=0.4\textwidth]{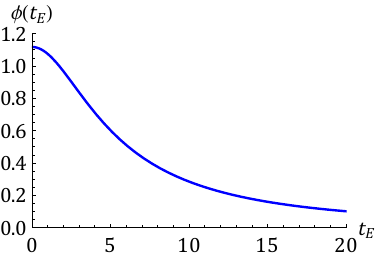}\\
		\includegraphics[width=0.4\textwidth]{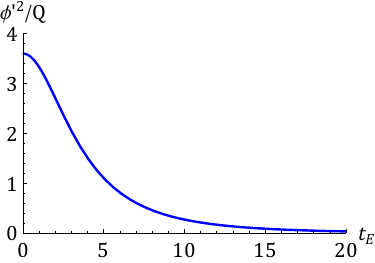}
		\includegraphics[width=0.4\textwidth]{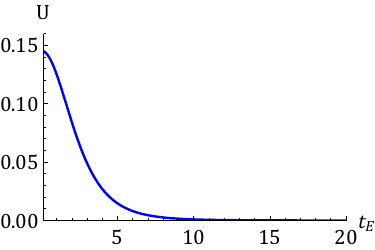}
		\centering
		\caption{\small Massless dilatonic wormhole with $\beta=1.$ In the upper row we show the evolution of the fields, and in the lower row one can see that the kinetic function $\dot\phi^2/Q$ as well as the fluctuation potential $U$ are positive definite. These solutions are thus perturbatively stable.}\label{fig:q}
	\end{figure}

	Massless dilatonic wormholes have $V(\phi)=0$ and $0 \leq \beta < \beta_c \equiv 2 \sqrt{\frac{2}{3}}.$ They are given by (e.g.,~\cite{Andriolo:2022rxc})
	\begin{align}
		\left\lbrace
		\begin{aligned}
		\mathrm{d}s^2 & = a_0^2 \cosh(2R) \left( \mathrm{d}R^2 + \mathrm{d}\Omega_3^2\right), \\
		\phi & = \frac{1}{\beta} \ln \left[ \frac{N^2}{3a_0^4}\cos^2 \left( \frac{\beta}{\beta_c} \arccos \frac{1}{\cosh(2R)}\right)\right], \\
		a_0^2 & = \frac{N}{\sqrt{3}} \cos\left( \frac{\pi}{2}\frac{\beta}{\beta_c}\right).
		\end{aligned}
		\right.
	\end{align}
	By plugging this solution into the expressions for $Q,$ $\phi^{\prime 2}/Q$ and $U$, we obtain the shapes shown in figure \ref{fig:q}. As expected from the expansion near the origin, the function $Q$ touches zero at the neck of the wormhole while the kinetic factor $\phi^{\prime 2}/Q$ and potential $U$ are regular there. Both the kinetic factor and the potential are regular and positive throughout, so that any fluctuation ${\cal R}$ will increase the Euclidean action ($S_\text{E}^{(2)}>0	$), see equation \eqref{eq:S2inphysicaltime}. Therefore these wormholes are stable.
	
	\subsection{Massive dilaton}

	\begin{figure}[h!]
		\begin{subfigure}{0.49\linewidth}
			\includegraphics[width=0.95\linewidth]{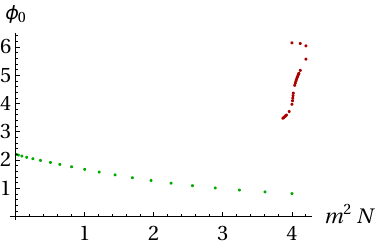}
			\subcaption{$\beta=\sqrt{2}$\label{subfig:sqrt2}}
		\end{subfigure}
		\begin{subfigure}{0.49\linewidth} 
			\includegraphics[width=0.95\linewidth]{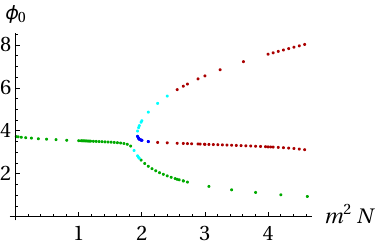}
			\subcaption{$\beta=1.579$} \label{figsub}
		\end{subfigure}\vspace{0.2cm}
		\begin{subfigure}{0.49\linewidth}
			\includegraphics[width=0.95\linewidth]{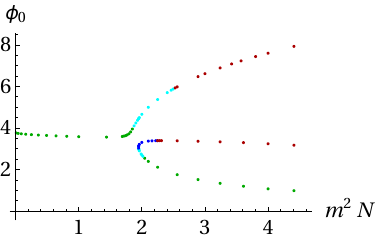}
			\subcaption{$\beta=1.58$}
		\end{subfigure}
		\begin{subfigure}{0.49\linewidth}
			\includegraphics[width=0.95\linewidth]{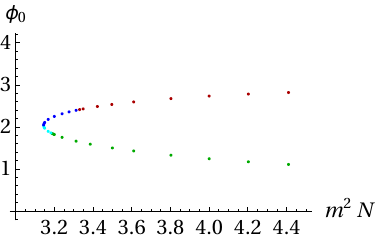}
			\subcaption{$\beta=1.64$}
		\end{subfigure}\caption{Summary plots for various values of the dilaton coupling $\beta.$ Dots indicate the existence of a wormhole solution, with the initial dilaton value $\phi_0(t_\text{E}=0)$ given as a function of the dilaton mass $m$ and axion charge $N.$ The colour coding is as follows: Green dots indicate stable solutions with $Q>0,\, U>0$ everywhere. Cyan dots indicate stable solutions with $Q>0$ everywhere, but $U<0$ in some regions. Blue dots indicate unstable solutions with one negative mode ($Q>0$ everywhere, $U<0$ in some regions). Red dots indicate solutions for which the kinetic function $Q$ contains negative regions, so that we cannot meaningfully study their stability in the present setup.\label{fig:betasummary}}
	\end{figure}

	We now turn to the physically more relevant case of the dilaton acquiring a mass. As discussed earlier, in this case many solutions exist that generalise the original Giddings-Strominger-type wormholes \cite{Andriolo:2022rxc,Jonas:2023ipa}. In fact, there are two dilaton coupling values that are most relevant for classifying solutions: the critical value $\beta_c=2\sqrt{\frac{2}{3}}\approx 1.63$ above which no massless solutions exist, and the ``inversion'' value $\beta_i,$ with $1.579 < \beta_i < 1.580,$ above which the lowest $\phi_0$ solutions at large mass are no longer connected to those at low mass. In the next few paragraphs, we therefore present our results (summarised in figure \ref{fig:betasummary}) for the four representative dilaton couplings that were already illustrated in figure \ref{fig:bckgsol}: $\beta=\sqrt{2}$ (corresponding to small dilaton coupling), $\beta=1.579$ (just below the inversion point), $\beta=1.58$ (just above the inversion point) and $\beta=1.64>\beta_c$ (large dilaton coupling, above the critical value). In all cases we used the weight function $w=\dot{\phi}^2/Q$.
	
\begin{figure}[h]
		\centering
		\includegraphics[width=0.4\textwidth]{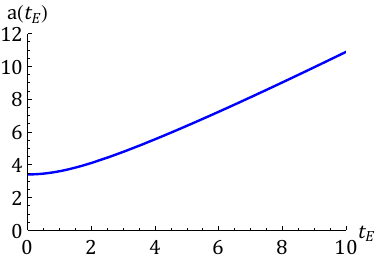}
		\includegraphics[width=0.4\textwidth]{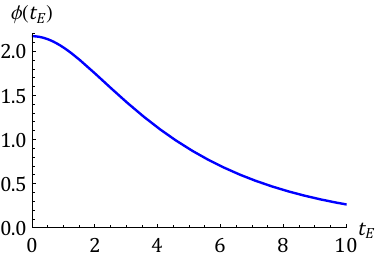}\\
		\includegraphics[width=0.4\textwidth]{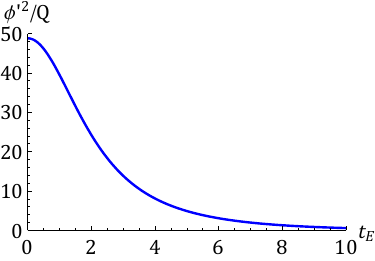}
		\includegraphics[width=0.4\textwidth]{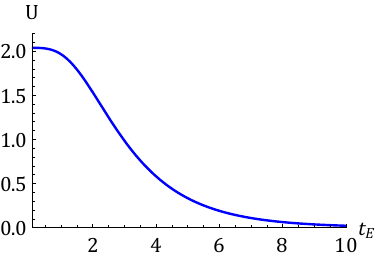}
		\centering
		\caption{\small Massive dilatonic wormhole with $\beta=1.579$ and $m^2 N = 2.2.$ This wormhole starts out at $\phi_0 \approx 2.1705146359.$ In the upper row we show the evolution of the fields, and in the lower row the kinetic function $\dot\phi^2/Q$ as well as the fluctuation potential $U.$ Both are positive definite, and thus this is an example of a perturbatively stable massive axion-dilaton wormhole.}\label{fig:stable}
	\end{figure}

	In the four cases, the solution with the smallest potential (i.e.  lowest $\phi_0$ value) is always stable, and represents the dominant solution, with the smallest Euclidean action. For $\beta<\beta_i,$ the dominant solutions are smoothly connected to the massless limit (as for $\beta=\sqrt{2}, 1.579$), while for $\beta>\beta_i$ the dominant solutions manifest a jump to the new bifurcating branch wherever the new branch exists (as for $\beta=1.58, 1.64$). Most of the solutions with lowest $\phi_0$ have positive definite kinetic and potential terms throughout, so that their stability is manifest. Those solutions are indicated by green dots in figure \ref{fig:betasummary}. Typical forms of their kinetic and potential functions are shown in figure \ref{fig:stable} for illustration. However, some of the lowest $\phi_0$ solutions have regions where the fluctuation potential $U$ turns negative. In those cases, we employ the nodal theorem: we look at the number of nodes of the zero eigenvalue $\lambda=0$ solution. If it has no node, then the wormhole is stable. By contrast, if it has a node, then a (nodeless) eigenmode with lower, i.e.,~negative, $\lambda$ must exist\footnote{It may help the intuition to think about the lowest level wave functions of a particle in a box (an infinite potential well) of length $L$. These are given by $\psi(x) \propto \sin\left( \frac{n\pi}{L}x\right)$ with $n \in \mathbb{N}^*$ and have energy levels $E_n=\frac{n^2\pi^2\hbar^2}{2mL^2}.$ More nodes in $\psi$ correlate with higher $E_n.$}. For the lowest $\phi_0$ solutions in which the potential $U$ contains a negative region, we find that no negative mode exists and that these solutions are perturbatively stable. Similar wormhole solutions without negative mode also exist on parts of the new bifurcating branches of solutions, see the plots for $\beta=1.579,\,1.58,\,1.64.$ These perturbatively stable solutions are all indicated by cyan dots in the figures. 
	
\begin{figure}[h]
		\centering
		\includegraphics[width=0.4\textwidth]{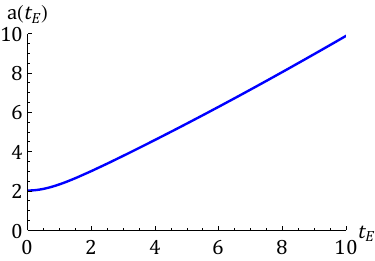}
		\includegraphics[width=0.4\textwidth]{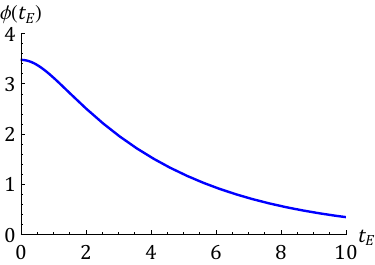}\\
		\includegraphics[width=0.4\textwidth]{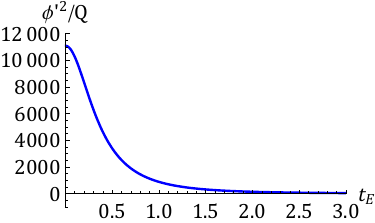}
		\includegraphics[width=0.4\textwidth]{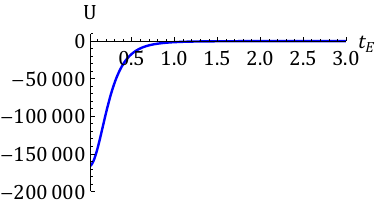}
		\centering
		\caption{\small Massive dilatonic wormhole with $\beta=1.579$ and $m^2 N = 2.2.$ This wormhole starts out at $\phi_0 \approx 3.4770669337.$ In the upper row we show the evolution of the fields, and in the lower row the kinetic function $\dot\phi^2/Q$ as well as the fluctuation potential $U,$ which starts out negative. This solution turns out to admit a negative mode, shown in the left panel of figure \ref{fig:negmode}, and is thus unstable.}\label{fig:mass}
	\end{figure}

\begin{figure}[h!]
		\centering
		\includegraphics[width=0.44\textwidth]{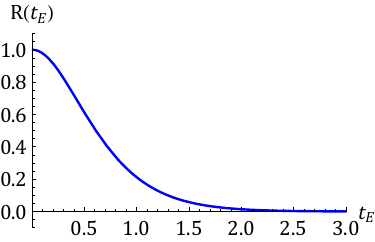}
		\includegraphics[width=0.44\textwidth]{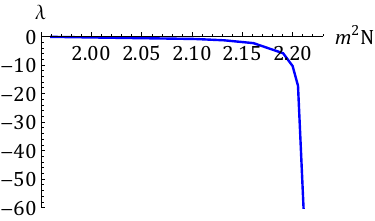}
		\caption{\small Left panel: the negative mode associated with the wormhole solution at $\beta=1.579, m^2N=2.2$ shown in figure \ref{fig:mass}. The eigenvalue is $\lambda \approx -10.3602531840.$ One can see that the eigenmode is concentrated near the neck of the wormhole, in the region where the fluctuation potential is negative. Right panel: negative eigenvalues as a function of $m^2N,$ for $\beta=1.579$ (for the blue dot solutions near the bifurcation point shown in figure \ref{figsub}) . The eigenvalues diverge near $m^2N \approx 2.22.$ }\label{fig:negmode}
	\end{figure}

\begin{figure}[h!]
		\centering
		\includegraphics[width=0.31\textwidth]{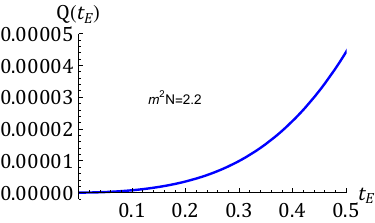}
		\includegraphics[width=0.31\textwidth]{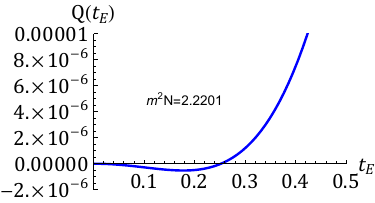}
		\includegraphics[width=0.31\textwidth]{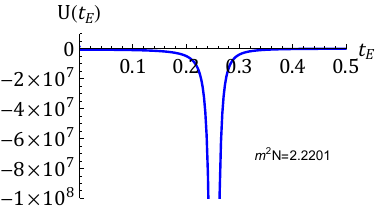}
		\caption{\small A closer look at the intermediate-$\phi_0$ solutions for $\beta=1.579$ shows that the kinetic function $Q$ evolves from starting out positive to starting out negative as the dilaton mass and axion charge are increased past $m^2N \approx 2.22.$ This is the reason that the eigenvalues of the negative modes of these solutions (blue dots in figure \ref{figsub}) diverge here, as shown in the right panel of figure \ref{fig:negmode}. When $Q$ starts out negative and then crosses zero, this also brings about a singularity in the fluctuation potential $U,$ as shown in the right panel above.}\label{fig:negQ}
	\end{figure}

	On the bifurcating branches for $\beta=1.579,\, 1.58,\, 1.64,$ leaving the bifurcation point in the other direction than that of the stable solutions just described, we find that a negative mode develops. These solutions are thus unstable, and their action is higher than that of the solutions on the opposite side of the bifurcating point (at the same values of $m^2 N$). We indicate such unstable solutions by blue dots in the figures. For illustration, we show an example at $\beta=1.579$ and $m^2N=2.2$ in figure \ref{fig:mass}. The figure shows the kinetic function $\dot\phi^2/Q,$ as well as the fluctuation potential $U$ which starts out negative but then turns (slightly) positive. This solution admits a negative mode in its perturbation spectrum, shown in the left panel of figure \ref{fig:negmode}. The negative mode is concentrated in the region where the effective potential is negative. If we plot the eigenvalues for different values of $m^2 N,$ as in the right panel of figure \ref{fig:negmode}, then we notice that the eigenvalues become more and more negative and diverge to $-\infty$ near $m^2N \approx 2.22.$ Although the precise numerical values of the eigenvalues depend on the measure chosen, we have verified that for different choices of measure (we tried $w=\dot\phi^2, \dot\phi^2H^2,\dot\phi^2/H^2$) the divergence always occurs. This divergence can be understood to be a consequence of the fact that the kinetic function $Q$ suddenly starts out negative for $m^2N \gtrapprox 2.22,$ see figure \ref{fig:negQ}. We indicate solutions that contain regions with negative $Q$ with red dots in the figure. We will comment further on these cases below.
	
	For $\beta>\beta_c$ (in our example $\beta=1.64$), we obtain the same results as for the bifurcating branches of the $\beta=1.58$ case. In other words, from the bifurcation points there are solutions that are stable at low $\phi_0,$ while the solution emanating at larger $\phi_0$ contain a negative mode. At sufficiently large $m^2N$ the kinetic function $Q$ starts to contain regions where it goes negative, and where we cannot assess the stability of the solutions reliably.
	
	For $\beta=\sqrt{2}$, solutions with low $\phi_0$ are stable, and they connect smoothly to the massless case as $m^2N \to 0.$ However, we find that the entire new branch contains regions where $Q<0.$ These are again marked with red dots in figure \ref{subfig:sqrt2}.

The cases where the kinetic function $Q$ contains negative regions warrants further comments. At least naively, a negative kinetic term indicates the presence of a ghost, and thus an infinite number of negative modes. What is more, when $Q$ crosses zero, the effective potential develops a singularity -- see figure \ref{fig:negQ} for an illustration. That said, let us note however that the quadratic action \eqref{quad} resembles that derived for the analysis of metastable vacuum decay in gravity coupled to a scalar field \cite{Lavrelashvili:1985vn}. There it was also noticed that the corresponding factor $Q$  can become negative for some Euclidean bounce solutions. As just mentioned, naively this leads to a catastrophic instability (infinitely many negative modes) and is known as the negative mode problem \cite{Tanaka:1992zw}. In spite of much work in this direction by different research groups
\cite{Lavrelashvili:1985vn,Tanaka:1992zw,Lavrelashvili:1998dt,Khvedelidze:2000cp,Lavrelashvili:1999sr,Gratton:2000fj,Lavrelashvili:2006cv,Dunne:2006bt,Battarra:2012vu,Battarra:2013rba,Lee:2014uza,Koehn:2015hga,Bramberger:2019mkv,Jinno:2020zzs}
no fully satisfactory solution of this problem has been found. See \cite{Lavrelashvili:1999sr,Jinno:2020zzs} for comparative analyses of different approaches. When the factor $Q$ is positive it was found \cite{Khvedelidze:2000cp,Gratton:2000fj} that Coleman-De Luccia bounces \cite{Coleman:1980aw}
have exactly one negative mode in their linear perturbation spectrum, which makes the decay picture coherent. 
It was later shown in \cite{Koehn:2015hga} that when $Q$ becomes negative along the bounce, the tunneling negative mode
continues to exist, while on top an infinite tower of negative modes appears with support in the region where $Q<0$. 
Typically these negative modes are well separated \cite{Lee:2014uza,Koehn:2015hga}, namely the frequency of the tunneling negative mode is much lower than the frequency of negative modes due to negativeness of $Q$. 
In the present analysis we are facing the same problem, but in a somewhat more severe form: for some solutions $Q$ is already negative at the neck of the wormhole, rendering the eigenvalue equation untractable. Also, as shown in Appendix~\ref{app:hamiltonian}, if we use a Hamiltonian treatment instead, the kinetic factor itself can become singular along the wormhole solution. Nevertheless, it remains unclear if this represents a physical, rather than merely a mathematical, problem. This is because the exact location where a problem appears depends on the variables used: for instance a canonical transformation can shift the location where the kinetic function goes negative or becomes singular. Thus there remains the hope that if one were to find suitable variables, such cases could also be analysed consistently. We leave the search for such variables for future work.


		

\subsection{Expanding wormholes}

	Expanding wormhole solutions are solutions whose scale factor is initially contracting (i.e., $\ddot{a}|_0<0$) so that, when analytically continued to Lorentzian time, they are in fact expanding. These solutions can mediate the nucleation of expanding baby universes \cite{Lavrelashvili:1988un}, hence the name. In~\cite{Jonas:2023ipa} we found numerous such solutions numerically both for axion-gravity coupled to a massive dilaton, and coupled to a scalar field with a double well potential. We immediately see that we cannot analyse the stability of these solutions using the above formulation because those expanding wormholes necessarily contain a minimum for the scale factor, away from $t_\text{E}=0,$ where $\dot{a}=0$. At that point, the function $Q=H^2-\kappa^2\dot{\phi^2}/6$ will be negative, and this will occur for all expanding wormhole solutions. Hence we do not obtain a regular Sturm-Liouville problem for these solutions, and cannot draw conclusions about their stability. As described above, we must leave their analysis to future work, and a search for better fluctuation variables.

\section{Conclusions} \label{sec:conclusions}

In the present research we have studied the perturbative stability of a wide class of axion-dilaton wormholes, both for massless and massive dilatons. The wormholes under consideration are Euclidean, with non-zero axionic charge, and contain asymptotically flat regions. Our main finding is that there exist stable wormhole solutions of this type at all values of the dilatonic mass~$m$ (including the massless case) and axionic charge~$N$ that we analysed. The dominant solutions are those that have small values of the dilaton potential. 

However, wormhole solutions at larger values of the dilaton potential also exist at sufficiently large~$m^2N.$ These solutions are characterised by more involved field configurations, containing oscillations in both the scalar field and the scale factor. Some of these are also stable (though sub-dominant), while others admit a negative mode in their fluctuation spectrum. In particular, when new branches of solutions come into existence at bifurcation points (as the parameter combination~$m^2N$ is increased), an additional zero mode appears at the bifurcation point. On one side of the new branch this zero mode turns into a stable fluctuation, while on the other side it becomes a negative mode signalling a perturbative instability. 

At all values of the dilatonic coupling $\beta$ we noticed that some solutions develop regions in which the kinetic term $Q$ for the fluctuations becomes negative. This is a manifestation of the long known negative mode problem \cite{Lavrelashvili:1985vn} in the present context.  In spite of many efforts by different groups in the last almost 40 years \cite{Lavrelashvili:1985vn,Tanaka:1992zw,Lavrelashvili:1998dt,Khvedelidze:2000cp,Lavrelashvili:1999sr,Gratton:2000fj,Lavrelashvili:2006cv,Dunne:2006bt,Battarra:2012vu,Battarra:2013rba,Lee:2014uza,Koehn:2015hga,Bramberger:2019mkv,Jinno:2020zzs}, a fully satisfactory solution of this problem has not been found yet. We just note that the precise parameter region where this problem occurs depends on the choice of perturbation variables used (and it also differs markedly depending on whether one uses a Lagrangian or Hamiltonian formalism), suggesting that the negativity of $Q$ is a mathematical obstacle rather than a physical instability. What is more, in the context of Coleman-DeLucchia instantons it was noted in \cite{Koehn:2015hga} that when a parameter is varied such that $Q$ passes through zero, the existing negative mode continues to exist (and its eigenvalue evolves continuously), while an additional infinite tower of separate negative modes appears. This suggests that the additional tower of modes may be unphysical, though the continuously existing negative mode certainly seems to be associated with a physical instability. In the present context, the singularities associated with $Q$ passing through zero were stronger, and prevented us from carrying out a similar analysis while varying background parameters. However, a way of overcoming these singularities might conceivably exist and if so would offer valuable clues as to the number of physical negative modes. That said, one can think of several ways that might allow one to overcome the problem with negative $Q$ altogether: using canonical transformations in order to find proper variables in which $Q$ is positive throughout the wormhole, or combining different variables with the aim of covering the whole wormhole with positive $Q$ patches. One might also be able to extend the model with extra matter fields and see if it is possible to find better defined variables in such an extended theory. Certainly, more work will be required to understand the stability in these cases.

We should point out that we focussed our analysis on the homogeneous sector of the perturbations. This is the sector that we expect to be the most relevant for the stability of the solutions, as higher harmonics come with additional gradient energy, which always increases the energy. In fact, for the case of purely axionic wormholes (without dilaton), this is precisely what was found in \cite{Loges:2022nuw}. By contrast, the instabilities that we found (i.e.,~the negative modes associated with some of the wormholes) stem from homogeneous perturbations of the dilaton-gravity system, with the axion remaining frozen. Thus we can identify the dilaton as the actual source of the instabilities we found. This is significant, as in string theory one always expects an axion to be paired with a dilaton field.

In fact, if one tries to connect string theoretic solutions to phenomenology, it is important that the dilaton obtain a mass. This is because the dilaton contributes to determining the couplings between gravity and matter fields, and observationally the coupling ``constants'' are indeed found to be constant, both over cosmological time and length scales \cite{Uzan:2002vq}. This means that our results apply to a potentially realistic context, in which the puzzles associated with Euclidean wormholes are sharpened.

To conclude, we have found that Euclidean wormhole solutions exist and are perturbatively stable, in the context of axion-dilaton-gravity, which is a framework that provides a realistic extension of know physics. This means that they will indeed contribute to gravitational path integrals and that the paradoxes associated with factorisation in holography, the tension with certain quantum gravity principles, the apparent loss of unitarity or the random values of coupling constants cannot be discarded that easily. On the contrary, our work suggests that these issues will have to be faced head on in future studies.

\acknowledgments
C.J.~thanks Simon Maenaut for useful discussions. C.J.~is supported by the postdoctoral junior fellowship 1227524N from the Research Foundation - Flanders (FWO). The work of G.L. is supported in part by the Shota Rustaveli National Science Foundation of Georgia with Grant N FR-21-860. J.L.L. gratefully acknowledges the support of the European Research Council (ERC) in the form of the ERC Consolidator Grant CoG 772295 ``Qosmology''. 	
	
\appendix
	\section{Hamiltonian approach}\label{app:hamiltonian}
	
In this Appendix, we will consider the Hamiltonian formalism for obtaining an unconstrained quadratic action. We will only sketch the calculation, starting again from equation \eqref{eqac}. Once more  we will work in the homogeneous sector, where we can set $S=0=W.$ Then the canonical Hamiltonian is given by \begin{align}
		\textbf{ H}  = & \Pi_\Psi \Psi' + \Pi_\Phi \Phi' - {\cal L} \nonumber \\= & -\frac{\kappa^2}{12a^2\sqrt{\gamma}}\, \Pi_\Psi^2 +\frac{\kappa^2}{2a^2\sqrt{\gamma}}\, \Pi_\Phi^2 + \frac{\kappa^2}{2}\phi' \, \Pi_\Psi \Phi \nonumber \\ & + a^2 \sqrt{\gamma} \left[ (\frac{3}{\kappa^2}  + \frac{9 N^2}{a^4}e^{-\beta\phi\kappa}) \Psi^2 + (\frac{1}{2}a^2 V_{,\phi\phi} -\frac{3\kappa^2}{4}\phi'^2 + \frac{\beta^2\kappa^2}{2}\frac{N^2}{a^4}e^{-\beta\phi\kappa}) \Phi^2 \right. \nonumber \\ & \left. \qquad \qquad  - 6\beta\kappa\frac{N^2}{a^4}e^{-\beta\phi\kappa} \Psi\Phi \right] \nonumber \\ & +A \left[-\frac{a'}{a}\Pi_\Psi +\phi' \Pi_\Phi \right. \nonumber \\ & \left. \qquad+ a^2 \sqrt{\gamma} \left( (-\frac{6}{\kappa^2}+6\frac{N^2}{a^4}e^{-\beta\phi\kappa}) \Psi +(3\frac{a'}{a}\phi' + a^2 V_{,\phi} - \beta \kappa \frac{N^2}{a^4}e^{-\beta\phi \kappa})\Phi \right)\right].
	\end{align}
	It follows that the linearised constraint is given by
	\begin{align}
		0 = & -\frac{a'}{a}\Pi_\Psi +\phi' \Pi_\Phi  \nonumber \\ &  + a^2 \sqrt{\gamma} \left( (-\frac{6}{\kappa^2}+6\frac{N^2}{a^4}e^{-\beta\phi\kappa}) \Psi +(3\frac{a'}{a}\phi' + a^2 V_{,\phi} - \beta \kappa \frac{N^2}{a^4}e^{-\beta\phi \kappa})\Phi  \right)\,.
	\end{align}

	The time  evolution of variables involves the commutator with the Hamiltonian, 
	\begin{align}
		f' = \frac{\partial f}{\partial \eta} + [f,\textbf{ H}] = \frac{\partial f}{\partial \eta} + \sum_{q=A,\Psi,\Phi} \left(\frac{\partial f}{\partial q}\frac{\partial \textbf{ H}}{\partial \Pi_q} - \frac{\partial f}{\partial \Pi_q}\frac{\partial \textbf{ H}}{\partial q}\right),
	\end{align}
	where in $ \frac{\partial f}{\partial \eta}$ the time derivative acts on explicit time-dependent coefficients, not on the variables themselves.
	
	There is a primary constraint, due to the vanishing of the momentum associated with the lapse, $\Pi_A=0,$ which we denote by
	\begin{align}
		{\cal C}_1 = \Pi_A = 0\,.
	\end{align}
	For consistency its time derivative must also vanish,
	\begin{align}
		{\cal C}_2 & \equiv {\cal C}'_1 = -\frac{\partial \textbf{ H}}{\partial A} \nonumber \\ & = \frac{a'}{a}\Pi_\Psi -\phi' \Pi_\Phi  + a^2 \sqrt{\gamma} \left( (\frac{6}{\kappa^2}-6\frac{N^2}{a^4}e^{-\beta\phi\kappa}) \Psi +(-3\frac{a'}{a}\phi' - a^2 V_{,\phi} + \beta \kappa \frac{N^2}{a^4}e^{-\beta\phi \kappa})\Phi \right).
	\end{align}
	Thus consistency of the imposition of the primary constraint gives a secondary constraint that requires the linearised Friedman equation to be satisfied. Its own time derivative must also vanish, and we find
	\begin{align}
		{\cal C}'_2 = \frac{a'}{a}{\cal C}_2\,.
	\end{align}
	This means that setting ${\cal C}_2$ to zero is in fact consistent, and does not give rise to a tertiary constraint.
	
	The constraints indicate that gauge degrees of freedom are present, and thus we should reduce the system to a physical degree of freedom. Here there are various choices. Obtaining a sensible limit when gravity is turned off suggests that we should not eliminate $\Phi$ but rather should try to eliminate variables having to do with $\Psi.$
			
	A useful choice is provided by the gauge-invariant variables (with $\kappa^2=1$ from here on)
	\begin{align}
		\Phi_g \equiv \Phi - \frac{\phi'}{6a^2\sqrt{\gamma}\left(1-\frac{N^2}{a^4}e^{-\beta\phi}\right)} \Pi_\Psi\,, \quad \Psi_g= \Psi + \frac{H}{6a^2\sqrt{\gamma}\left(1-\frac{N^2}{a^4}e^{-\beta\phi}\right)}\Pi_\Psi\,,
	\end{align}
	The path integral over $A$ then imposes the constraint ${\cal C}_2=0,$ which one can use to eliminate $\Pi_\Phi.$
	All terms involving $\Pi_\psi$ are now automatically incorporated in the field redefinition, and do not appear explicitly in the action. What is more, $\Psi_g$ appears without derivatives and can hence be integrated out, as it simply gives rise to a Gaussian integral which will yield a prefactor. This integration can straightforwardly be done by completing the square. In the end we are left with $\Phi_g$ only, with action
	\begin{align}
		S^{(2)}
		= \frac{1}{2}\int d^4x \sqrt{\gamma}  & a^3 \left(\frac{1}{{Q_\text{H}}}\dot{\Phi}_g^2  + U_{\Phi}{\Phi}_g^2 \right)\,,
	\end{align}
	where the kinetic term in the Hamiltonian approach is governed by a generalised $Q$ function
	\begin{align}
		\textrm{Euclidean} \qquad Q_\text{H}=1- \frac{a^2 \dot\phi^2}{6}  \frac{\left( 1+3\frac{N^2}{a^4}e^{-\beta\phi}\right)}{ (1-\frac{N^2}{a^4}e^{-\beta\phi})^2}
	\end{align}
	while the fluctuation potential is given by
	\begin{align}
		U_\Phi
		& = \frac{1}{Q_\text{H}}\left[ V^{\prime\prime} + \frac{2 \dot\phi^2}{\left( 1-\frac{N^2}{a^4}e^{-\beta\phi}\right)} + \beta^2 \frac{N^2}{a^6}e^{-\beta\phi} + \beta \frac{H \dot\phi N^2}{a^6} e^{-\beta\phi}\frac{\left( 1+3\frac{N^2}{a^4}\e^{-\beta\phi}\right)}{\left( 1-\frac{N^2}{a^4}\e^{-\beta\phi}\right)^2} \right] \nonumber \\ & \quad + \frac{\dot{Q}_\text{H}}{Q_\text{H}^2}\left( 3H-\frac{\dot{V}}{\dot\phi^2} +\frac{\beta N^2}{a^6 \dot\phi}\e^{-\beta\phi}\right)\,.
	\end{align}
	A prime denotes a scalar field derivative here, while a dot denotes a time derivative. The associated fluctuation equation (with weight function $\sqrt{\gamma} a^3$) reads
	\begin{align}
		-\frac{1}{a^3}\frac{\dd}{\dd t_\text{E}} \left( \frac{a^3}{Q_\text{H}} \frac{\dd\Phi_g}{\dd t_\text{E}}\right) + U_\Phi \Phi_g = \lambda\Phi_g \,.
	\end{align}
	
The equation of motion implies that at $t_\text{E}=0$ we have $a\ddot{a} + 1 - \frac{N^2}{a^4}\e^{-\beta\phi}\mid_{t_\text{E}=0}=0,$ and this relation implies that   $1 - \frac{N^2}{a^4}\e^{-\beta\phi}$ will start out negative (for GS-type wormholes) and reach unity asymptotically. Hence it will pass through zero at least once, leading to one or more singularities in $Q_\text{H}$ in all cases. Unfortunately, these singularities prevent one from using the Hamiltonian formalism in the study of these wormholes. We have also explored the possibility of performing various canonical transformations of the variables, but in each case one simply ends up shifting the singularities from one place to another. This unfortunately seems to limit the usefulness of the Hamiltonian method in the present setting, and motivates the use of the Lagrangian formalism in the main part of the paper.

\bibliographystyle{utphys}
\bibliography{biblio}

\end{document}